\documentclass{elsart}
\usepackage[authoryear]{natbib}
\usepackage{graphicx}
\usepackage{amssymb}
\usepackage{amsmath}
\usepackage{url}
\usepackage{lineno}

\begin{document}
\begin{frontmatter}
\title{Speciation-rate dependence in species-area relationships}
\author[NBI,CO]{Simone Pigolotti}
\and
\author[SMC]{Massimo Cencini}
\address[NBI]{The Niels Bohr International Academy,
  The Niels Bohr Institut, Blegdamsvej 17, DK-2100 Copenhagen,
  Denmark}
\address[CO]{Corresponding author, email: pigo@nbi.dk, Ph. +4535325238, 
Fax +4535325425}
\address[SMC]{INFM-CNR, SMC Center for Statistical
    Mechanics and Complexity, Dipartimento di Fisica Universit\`a di
    Roma ``La Sapienza'' Piazzale A. Moro, 2 00185 Rome, Italy and
    ISC-CNR, Via dei Taurini, 19 00185 Rome, Italy.}
\begin{abstract}

  The general tendency for species number (S) to increase with sampled
  area (A) constitutes one of the most robust empirical laws of
  ecology, quantified by species-area relationships (SAR).  In many
  ecosystems, SAR curves display a power-law dependence, $S\propto
  A^z$.  The exponent $z$ is always less than one but shows
  significant variation in different ecosystems.  We study the
  multitype voter model as one of the simplest models able to
  reproduce SAR similar to those observed in real ecosystems in terms
  of basic ecological processes such as birth, dispersal and
  speciation. Within the model, the species-area exponent $z$ depends
  on the dimensionless speciation rate $\nu$, even though the detailed
  dependence is still matter of controversy.  We present extensive
  numerical simulations in a broad range of speciation rates from $\nu
  =10^{-3}$ down to $\nu = 10^{-11}$, where the model reproduces
  values of the exponent observed in nature.  In particular, we show
  that the inverse of the species-area exponent linearly depends on
  the logarithm of $\nu$.  Further, we compare the model outcomes with
  field data collected from previous studies, for which we separate
  the effect of the speciation rate from that of the different species
  lifespans. We find a good linear relationship between inverse
  exponents and logarithm of species lifespans. However, the slope
  sets bounds on the speciation rates that can hardly be justified on
  evolutionary basis, suggesting that additional effects should be
  taken into account to consistently interpret the observed exponents.
\end{abstract}

\begin{keyword}
Species-area exponents \sep Spatial models \sep Dispersal \sep
  Voter model \sep Biodiversity \sep Neutral Theory
\end{keyword}

\end{frontmatter}

\section{Introduction}

Species-area relationships (SAR) quantify ecosystem richness and, in
particular, the spatial variations of biodiversity. These curves
measure the average number of species ($S$) present in a sample area
($A$) of a given ecosystem and usually display a triphasic shape
\citep{preston1960,rosenzweig1995,hubbell2001}. For small areas (below
the dispersal range) and large areas (continental scale), the number
of species rapidly increases with the area; while for
intermediate areas a slower, sub-linear growth is observed.  The
intermediate range is the most intriguing one and has gathered much
attention since its discovery.  Although many functinoal forms
have been proposed to fit the data in this intermediate regime
\citep{he1996,tjorve2003}, the most common and widely accepted ones
are the algebraic law $S= C A^z$ (with $z<1$ and $C$ a positive
constant) proposed by \citet{arrhenius1921} (see also
\citet{gleason1922}), and the logarithmic one $S\approx C\ln A$ due to
\citet{fisher1943}. A recent survey by \citet{drakare2006},
reconsidering most of the existing SAR studies from different
ecosystems, shows that the former provides a better fit in about half
of the cases. Even though any of the two hypothesis cannot be
\textit{a priori} discarded, much efforts across the years
\citep{preston1962,macarthur1967,connor1979,wright1988,kohn1994,durrett1996,hubbell2001,chave2002,he2002,martin2006}
have been devoted to explain the observed values of the exponent
$z$. Observations support the idea of a dependence of the exponent $z$
on quantities such as latitude \citep{allen2006} and body size of
considered species \citep{drakare2006}. Notwithstanding observations
and theoretical efforts, a satisfactory theory able to predict the
value of the exponent in different ecological situations is still
lacking.

On the theoretical side, two distinct viewpoints on ecosystems
organization correspond to different explanations for species-area
relationships. According to the first, larger areas contain a larger
variety of habitats and consequently can sustain a richer species
diversity \citep{kohn1994}. For the second viewpoint, species-area
relationships are the outcome of demographic processes such as
colonization, dispersal, speciation and local extinction, and do not
need environmental diversity for their
explanation~\citep{macarthur1967,hubbell2001}. We should mention a
third explanation, ascribing species-area relationships to statistical
biases induced by the skewedness of species abundance
distributions~\citep{he2002,martin2006}. As both niche-based and
neutral-dispersal mechanisms are able to sustain diversity, the hope
is to extract information on the importance of the different classes
of effects from the shape of SAR curves~\citep{chave2002}.

We consider the voter model as the simplest prototype of neutral
models able to generate non-trivial species-area relationships
\citep{durrett1996,hubbell2001,zillio2005,rosindell2007}. The model
accounts, in a simple way, for the processes of birth, local dispersal
and introduction of new species.  Its main parameter, $\nu$, is a
dimensionless number measuring the rate of appearance of new species
--- speciation events --- in units of the death rate.  The other
ingredient is the dispersal kernel, which quantifies the probability
for an individual of a species to colonize different locations in the
ecosystem. Speciation and dispersal are enough to produce triphasic
SAR curves resembling those observed in field data
\citep{chave2002,rosindell2007}. In particular, for local
(short-range) dispersal, the intermediate regime is well described by
a power-law behavior $S=C A^z$, with an exponent $z$ depending on the
speciation rate $\nu$. We mention that the logarithmic function is
reproduced by the voter model with global dispersal, when individuals
can invade all loci of the ecosystem \citep{coleman1981,chave2002}.
The logarithmic law is also retrieved, for large dispersal, for areas
being smaller than the dispersal range, indeed at these scales the
dispersal appears as if it was long range.  Understanding how the
exponent $z$ depends on the parameters of the model (in this case,
mostly on $\nu$) is fundamental to move a step toward the theoretical
prediction of the variations of experimentally observed exponents in
terms of ecological quantities.

However, this dependence has been source of some controversy in the
literature.  In a seminal paper, \citet{durrett1996} proposed a
formula according to which, in the limit small $\nu$, $z\!  \sim\!
1/\ln(1/\nu)$.  \citet{rosindell2007} suggest a power law relationship
between $z$ and $\nu$. Finally, the scaling argument of
\citet{zillio2005} predicts $z$ approaching a finite value $z\!\approx
\!0.2$ for vanishing $\nu$.  These discrepancies have not yet been
settled and, due to the weak dependence of $z$ on $\nu$, a clean
answer requires numerical simulations with $\nu$ varying over several
orders of magnitude. So far, only speciation rates $\nu\gtrsim
10^{-6}$ were explored as simulations at lower (possibly more
realistic) values of the speciation rate are computationally very
expensive.

In this paper we present results of simulations of the voter model
with speciation rates varying in a wide range of values from
$\nu=10^{-3}$ down to $\nu= 10^{-11}$, with the twofold aim of
disentangling the low speciation rate behavior and examine an
ecologically relevant range of parameters. Our findings are also
useful to assess whether neutral predictions are consistent with
realistic speciation rates \citep{hubbell2001}, a question which
raised a heated debate \citep{hubbell2003,ricklefs2003}.  SAR curves
resulting from our simulations are characterized by a power law
behavior with exponent $z$, displaying a logarithmic dependence on the
speciation rate and supporting \textit{de facto} Durrett-Levin's
scenario, even though with different numerical coefficients.  In
agreement with \citet{rosindell2007}, we also found that the exponent
$z$ is essentially insensitive to the dispersal range implying that,
accepting the hypothesis of the model, the observation of a
species-area exponent imposes strong constraints on the rate of
appearance of new species. In the Discussion section, we examine the
plausibility of the model predictions on the basis of data available
in the literature.  In particular, we consider the $z$-values reported
in the literature for different taxa and, due to the absence of
reliable data on speciation times, we study how measured exponents
depend on the lifespan, with the additional assumption that average
speciation times and lifespans are linked by a scaling relation. The
observed variations turn out to be much larger than those allowed in
the framework of the model; we finally discuss which effects may be
included to possibly achieve a quantitative description.

\section{Model}

We consider the voter model with mutation as defined by
\citet{durrett1996}. Individuals belonging to different species are
placed at each site of a two-dimensional $(L\!\times\! L)$-square
lattice and evolve according to the following dynamics. At each
time-step, a randomly chosen individual is killed, creating a gap
which is immediately filled, with probability $\nu$, by an individual
from a new species (not present in the ecosystem) --- speciation event
--- or, with a probability $1-\nu$, by a new individual of an already
existing (in the ecosystem) species chosen among those present in a
neighborhood (that will be detailed below) of the site ---
birth/dispersal event.  The dynamics is then advanced until the number
of species in the ecosystem reaches a statistically steady
value. Strictly speaking, the fact that empty locations are
immediately colonized means that the birth rate is infinite (see
discussion in \citet{durrett1996}). Therefore the basic time-step of
the dynamics correspond to a death event, and thus the dimensionless
parameter $\nu$ represents the speciation rate $\sigma$ measured in
unit of the death rate $d$. Equivalently, we can express $\nu$ as the
average species lifespan $t$ divided by the average time between
speciation events $t^{(s)}$ (we shall come back to this point in the
Discussion section).

As for the dispersal rule several options are possible.  The simplest
possibility is the nearest-neighbor rule, where the individual is
replaced by one of the species present in the four neighbor sites with
probability $1/4$. We will refer to this in the following as the
nearest-neighbor (NN) case. A more realistic choice is to use a
generic dispersal kernel introducing the probability $P(r)$ of a gap
being filled by a species whose representative individual is at a
distance $r$ from it. We adopt a computationally simple instance by
choosing the square kernel: we replace the individual with a copy of
another individual randomly chosen in a square of side $2K+1$ centered
on the gap. This choice does not represent a restriction as it has
been shown that the relevant quantity is the averaged square
dispersal distance and not the specific functional form of the kernel.
For instance, a Gaussian and a square dispersal kernel with the
  same squared dispersal distance produce very similar SAR
\citep{rosindell2007}.

We stress that, independently of the dispersal rule, the model is
completely neutral: all species (and individuals) undergo the same
dynamics, as differences among species arise only due to demographic
stochasticity.

Simulations have been efficiently performed by using the dual
representation of the voter model \citep{holley1975}, providing a way
to reconstruct the asymptotic configuration of the ecosystem by
tracing backward in time its evolution. An important advantage of the
dual representation is that it reconstructs the genealogy of each
individual up to the speciation event originating its species, meaning
that the system is ensured to have reached equilibrium.  Moreover, it
allows to implement open boundary conditions: the genealogy of an
individual can be reconstructed also when its ancestors are outside
the simulated area, which can thus be considered as a sample of a
virtually infinite ecosystem \citet{rosindell2007,rosindell2008}.
This means that we can interpret $\nu$ as a \textit{bona fide}
speciation rate, since immigration from outside the system is included
in the birth-dispersal process.  However, long-range immigration
events qualitatively different from local dispersal (i.e. seeds
transported by birds) can be modeled as an higher ``effective''
speciation rate $\nu$.  We managed to optimize the algorithm to
simulate the model for very low speciation rates, down to
$\nu=10^{-11}$. Details on how the simulations have been performed and
the statistics have been collected can be found in
Appendix~\ref{appa}; see also \citet{rosindell2008} for other possible
improvements of the coalescence algorithm.  As for the dispersal, we
explored both the NN and the square kernels, for the latter $K$ has
been varied in the range $K=3-64$, though we shall mostly present the
results for $K=7$ (see the discussion in the next section).

\section{Numerical results}

\begin{figure}[t!]
\begin{center}
  \includegraphics[width=0.6\textwidth]{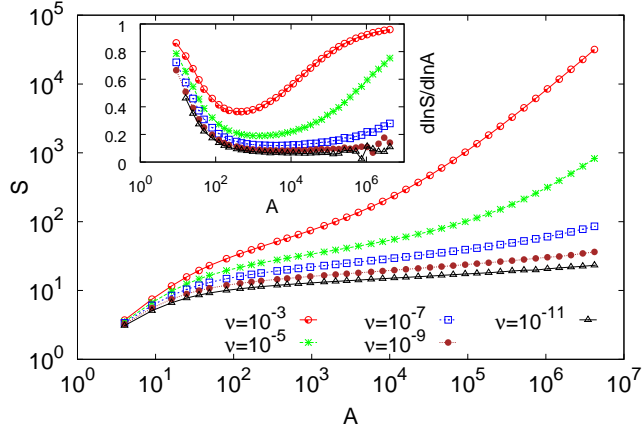}
\caption{Dependence of SAR curves on the speciation rate: $S$ vs $A$
 for different speciation rates $\nu\in [10^{-9}:10^{-3}]$ obtained
 with the square kernel with $K=7$ in a simulation sample of side
 $L=2048$, and averaging over $100$ independent realizations. Note the
 triphasic shape clearly observable for larger values of $\nu$. Inset:
 Logarithmic derivatives of the SAR curves, $\mathrm{d} (\ln S)
 /\mathrm{d} (\ln A$) vs $A$, a plateau identifies the intermediate
 regime and the plateau value the exponent $z$. Note that $z$
 increases with $\nu$ and the intermediate regime enlarges in width at
 decreasing $\nu$ up to invading almost all the simulation sample for
 small $\nu$ values.\label{figuraexample}
 }
\end{center}
\end{figure}

We begin studying SAR curves obtained at fixed dispersal range ($K=7$)
and varying $\nu$, as shown in Fig.~\ref{figuraexample}.  All
curves display a fast growth for small areas with a crossover, for
areas of the order of the dispersal kernel ($A\approx K^2$), to the
power-law regime. The final regime where the number of species becomes
linear with the area can be detected only for rather large
speciation-rate values, $10^{-5}\leq \nu\leq 10^{-3}$; to observe it
at lower values of $\nu$ much larger simulation samples would be
required. In the inset, we plot the ``local species-area exponent''
for each curve, $\mathrm{d}(\ln S)/\mathrm{d}(\ln A)$, which clearly
shows that the smaller $\nu$ the smaller the exponent becomes and the
larger is the range of scales where a well defined power-law behavior
establishes.  Finally, when the parameter $\nu$ is not too small, it
is possible to observe also the final linear regime which occurs for
areas much larger than $\nu^{-1}$ \citep{durrett1996}.

\begin{figure}[t!]
\begin{center}
  \includegraphics[width=1\textwidth]{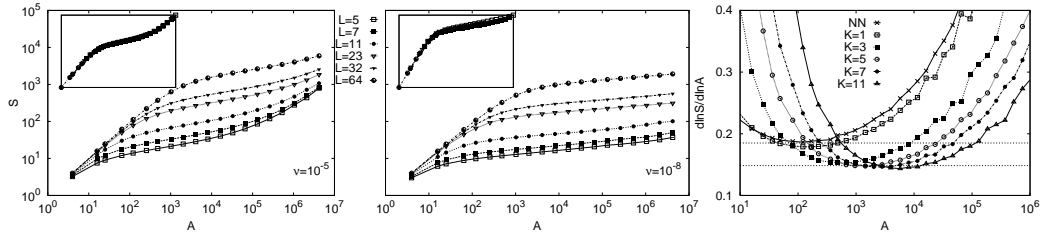}
 \caption{Dependence of SAR curves on the dispersal length. Left and
Middle panels: $S$ vs $A$, for $\nu=10^{-5}$ and $\nu=10^{-8}$
obtained with the square kernel with several values of $K$ in a
simulation sample with $L=2048$. Note that in the intermediate regime,
which starts and ends at different area values by changing $K$, the
slope of the curves is the same indicating that $z$ is independent of
$K$. Insets show the curves rescaled with the transformation
(\ref{eq:rescaling}).  Right panel: local slopes $\mathrm{d} (\ln S)
/\mathrm{d} (\ln A)$ vs $A$ for $\nu=10^{6}$ and different kernels: NN
and square with $K=1,3,5,7,11$. Note the tiny dependence of the
exponent (the region between the two horizontal lines) on the
dispersal length. For $K\gtrsim 5$ the plateau region (lower
horizontal line) does not change anymore apart from an horizontal
shifting of its onset towards larger areas.\label{figurakernel} }
\end{center}
\end{figure}

Figure~\ref{figurakernel} (left and middle panel) exemplifies the
behavior of species-area curves at fixed $\nu$ and different dispersal
range $K$.  At increasing the dispersal range the onset of the power
law regime shifts at larger areas, apparently without affecting the
exponent.  A more careful analysis of the local exponents $\mathrm{d}
(\ln S) /\mathrm{d} (\ln A)$, shown in the right panel, detects a
dependence of the value of the exponent on the dispersal range when
this is small, $K \lesssim 5$, including the NN case.

On the other hand, when $K\gtrsim 5$, we did not observe any
appreciable corrections to the value of the exponent. The independence
against variations of $K$, when it is large enough, has been
quantified by \citet{rosindell2007}, who have shown that curves
obtained with different (not too small) $K$ can be rescaled on a
universal function of $A$ and $\nu$ only via the transformation:
\begin{equation}
S=f(A,\nu,L)=K^r\phi(A/K^r,\nu)
\label{eq:rescaling}
\end{equation}
characterized by the scaling exponent $r\approx 1.97$. We checked that
this relation holds also with the small values of $\nu$ that we
studied, for instance the insets of Figure~\ref{figurakernel} (left
and middle panel) show it for $\nu=10^{-5}$ and $\nu=10^{-8}$.  We
will then study in the following the NN and the $K=7$ cases, the
former being that originally studied by \citet{durrett1996} and the
latter being representative of the behavior of the model for large
average dispersal distances.

We now turn to the main results of this paper about the dependence of
$z$ on $\nu$.  In Figure~\ref{figuraexponents} (left) we show the
exponent $z$ as a function of the speciation rate $\nu$ (see
Appendix~\ref{appa} for a discussion on how we estimated $z$).  We
observe a clear discrepancy with previous
predictions~\citep{durrett1996,rosindell2007} (also shown in the
picture). In particular, for $\nu\ll 1$, we found the data to
fall into a straight line when plotting $1/z$ vs. $\ln(\nu)$
(Fig.~\ref{figuraexponents} right), suggesting the following
functional dependence
\begin{equation}\label{ourfit}
z=\frac{1}{q+m\ln(\nu)}\,,
\end{equation}
by which we obtained a best fit to the data with $q\approx -3.3$ and
$m\approx -0.72$. To compare our results with previous studies of
these models, notice that the power-law fit suggested in by
\citet{rosindell2007} agrees with the data in the same range of
speciation rate values, i.e.  $\nu\ge10^{-5}$. Deviation from a power
law behavior are clearly observed for lower values of $\nu$, where the
data also rule out the saturation at $z\approx 0.2$ predicted by
\citet{zillio2005}. Actually, our fit confirms \citet{durrett1996}
prediction of a logarithmic decay of $z$ with $\nu$, up to corrections
order $\mathcal{O}(\ln (\ln( \nu)))$. However, the fitting parameters
$m$ and $q$ for both the square kernel with $K=7$ and the NN kernel
are very different from those of \citeauthor{durrett1996} (see caption
of Fig.~\ref{figuraexponents}). We conjecture that the differences in
prefactors could be caused by two different assumptions used by
\citeauthor{durrett1996} to derive the dependence of $z$ on $\nu$. The
first is about pre-asymptotic effects: the statistical results used by
Durrett and Levin are strictly valid only when $t\rightarrow \infty$
which requires $\nu\to 0$, while finite-time corrections may affect
the exponent value. In this respect, also for our data the $\nu \to 0$
limit seems to be crucial for the validity of the fit~\eqref{ourfit}.
 The second is the assumption that a power law
regime establishes from $A=1$ to $A=\nu^{-1}$. Conversely, we observe
the onset of the power law for areas being slightly larger than $1$
even in the NN case. Moreover, the crossover to the linear asymptotic
regime begins for areas quite smaller than $1/\nu$.

It should be noticed that discrepancies in the numerical factors have
profound implications when the model is used to estimate a speciation
rate from an observed species area exponent. The logarithmic
dependence of $z$ on $\nu$ makes, in fact, $\nu$ exponentially
dependent on $z$. We will discuss in the next section how this
dependence can be compared with experimental data.  It should also be
remarked that both \citeauthor{durrett1996} prediction and
equation~(\ref{ourfit}) are valid for small values of $\nu$ and can
lead to incorrect results, such as negative $z$, for $\nu$ close to
$1$.

\begin{figure}[t!]
\begin{center}
  \includegraphics[width=0.8\textwidth]{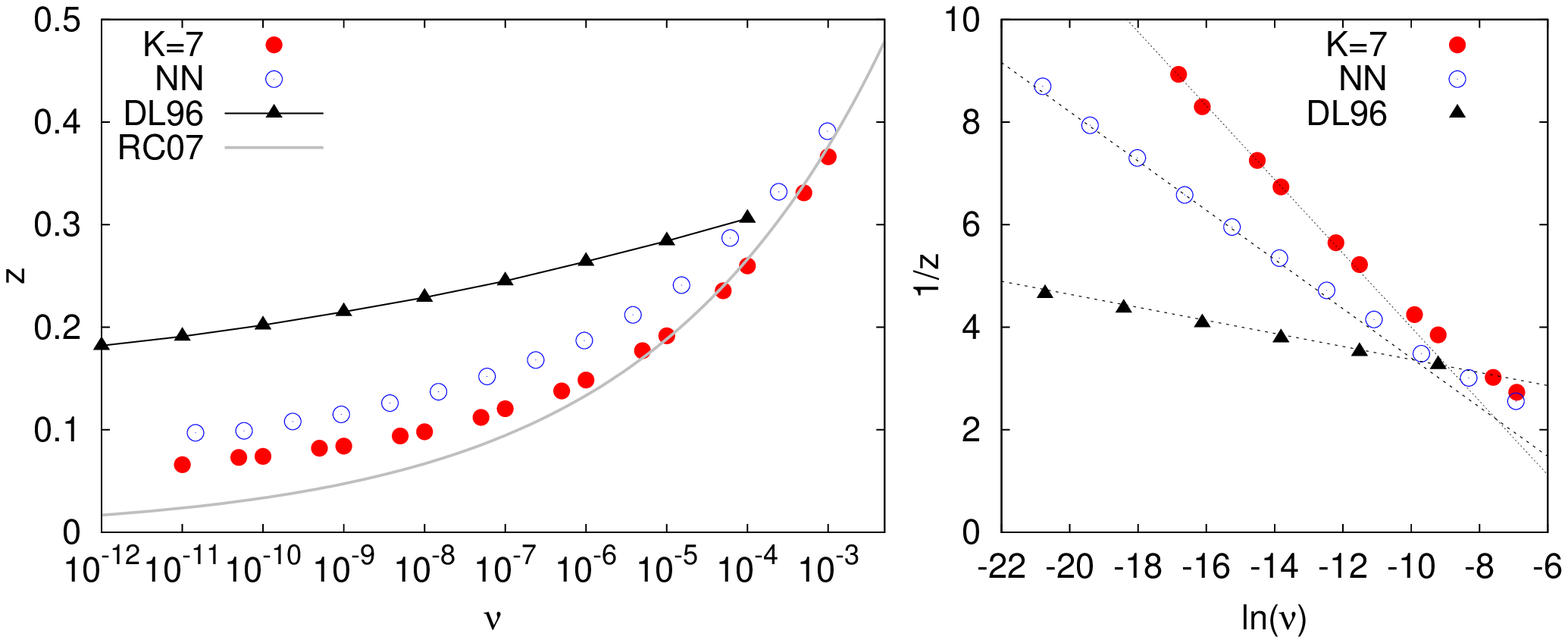}
\caption{Dependence of the species-area exponent on the speciation
rate.  (left) SAR exponent $z$ versus the speciation rate $\nu$ for
the NN and square kernel with $K=7$. For comparison it is also shown
the prediction by \citet{durrett1996} and the power-law dependence
suggested by \citet{rosindell2007}. (right) Same data as left but
shown plotting $1/z$ vs $\ln(\nu)$. The straight lines reports the
best fit with (\ref{ourfit}) with parameters: $m\!=\!-0.48\pm 0.02$,
$q\!=\!-1.4$ for NN data, $m\!=\!-0.72\pm 0.02$, $q\!=\!-3.2$ for
square kernel data ($K\!=\!7$) and $m\!=\!-0.127\pm 0.002$,
$q\!=\!2.10$ for Durrett-Levin tabulated data.  Errors on the
exponents are of the order of symbol sizes, see Appendix A for
details.}\label{figuraexponents} 
\end{center}
\end{figure}

\begin{figure}[t!]
\begin{center}
  \includegraphics[width=0.6\textwidth]{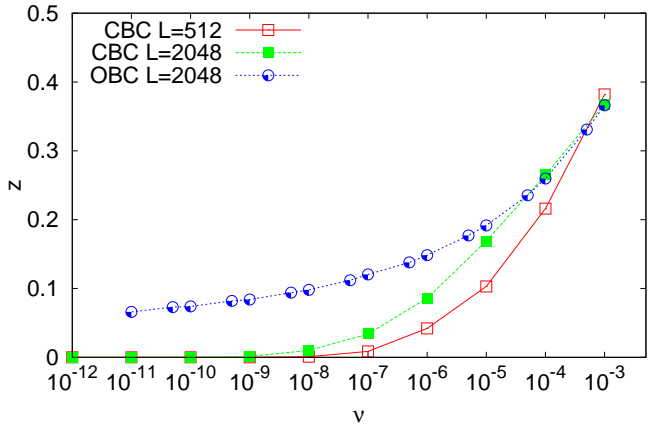}
\caption{$z$ vs $\nu$ for the square kernel with $K=7$ with closed boundary
conditions (CBC) at $L=512$ and $L=2048$ compared with the result with open
boundary conditions (OBC) and $L=2048$.  Note how the decrease of $z$ with $\nu$
becomes more and more dramatic at decreasing the
size.\label{fig:closed-boundary} }
\end{center}
\end{figure}

All simulations so far presented have been performed with open
boundary conditions, which are appropriate when the sample is a
homogeneous portion of a much larger ecosystem. However, closed
boundary conditions can be of interest for modeling confined
ecosystems such as islands.  Intuitively, open boundaries allow new
species to immigrate into the sampled system from the external
infinite ecosystem, independently of the speciation events. Closed
boundaries exclude this possibility and are thus expected to reduce
the exponent $z$ and, in general, species richness.  Fixing the
speciation rate $\nu$ the decreasing of $z$ becomes more and more
efficient as the system size decreases, and at fixed size the effect
is the stronger the smaller is $\nu$. The closed boundaries effects
becomes dramatic for simulation samples with $A\approx 1/\nu$, and, in
particular, we observed that for $\nu$ such that $A\lesssim 1/\nu$,
biodiversity is definitely lost, i.e. $z=0$ (see
Fig~\ref{fig:closed-boundary}). Notice that when the system size is
large (i.e. $A\ge 2048^2$) and the speciation rate is not too small
(i.e. $\nu\le 10^{-4}$) the exponent is essentially the same of the
open boundary case. We also remind that, if islands are modeled
by closed boundaries, the parameter $\nu$ should be meant to include
the immigration rate of new species \citep{macarthur1967}, since
dispersal from outside the system is forbidden in this case.

We stress here that the implementation of closed boundary is a very
simplistic way of describing confined ecosystems and more
sophisticated effects may be relevant in these cases.  For example, it
has been shown that the dynamics at the edge of ecosystems can be
quite different from that in the bulk \citep{laurance2006}. This means
that ``open'' situations, when the sample is part of a much larger and
homogeneous system, provide a much safer comparison between models and
field data.

\section{Comparison with empirical data and discussion}

Species-Area relationships have been subject of intense experimental
research in a variety of ecosystems, and the range of variability of
the exponent describing the intermediate regime goes from $z \approx
0.05$ in bacteria \citep{horner2004} to $z= 0.4-0.5$ in some plants
community (see \citet{drakare2006} for an exhaustive review of field
observations).  According to our results, the voter model with
speciation displays a variability of $z$ in the same range when the
speciation rate is allowed to vary over several orders of
magnitude. It is thus tempting to go in the direction of a more
quantitative comparison between field data and the model results.

As a first step, we assume that the model is able to describe the main
features of groups of trophically similar species and explore the
consequences of this assumption. This requires that a speciation rate
and a dispersal range for the whole group of species can be properly
defined, although we know that speciation rates \citep{mariakeva2004}
and dispersal ranges \citep{nathan2000,kinlan2003} may have
significant variations from species to species. However, in the model
the exponent is essentially independent on the dispersal range and
depends only logarithmically from the speciation rate, so that these
variations might be disregarded treating all species in the group as
having the same ``average'' dispersal range and speciation rate.

As far as dispersal is concerned, we only found a tiny dependence for
very short dispersal range, around $K<5$. Above these values the
exponent is independent of the dispersal range confirming previous
results \cite{rosindell2007}. Realistic average dispersal ranges
\citep{nathan2000,kinlan2003} are certainly far from the short
dispersal case, due to animal motility or wind for seeds. Therefore,
we assume that the dispersal range of real groups of species is always
in the range where the exponent is dispersal-independent.  It is
however worth remarking that the dispersal range can still affect the
spatial biodiversity via the power-law prefactor, whose increase can
lead to a large number of species that, when $z$ is small, increases
very slowly with the area. In this respect, the model outcomes are in
contrast with interpretations of low values for $z$ in bacteria as an
effect of large dispersal distances as argued in \citet{drakare2006}
and \citet{horner2004}.

What about speciation?  Unfortunately, we do not have ecological data
allowing us to directly estimate the frequency of speciation
events. Data from fossils suggest an average speciation rate on Earth
of about three specie per year \citep{sepkoski1998}, but it is hard to
infer from this number a reasonable rate for a living system. Also
estimates based on mutation rates \citep{mariakeva2004} could be
flawed due to genetic bottlenecks and phenomena like horizontal gene
transfer \citep{jain1999}. Moreover, as discussed in the Model
section, the parameter $\nu$ should be interpreted as an ``effective''
speciation rate, incorporating also long-range dispersal events.
Within the model framework, our results show that species-area
exponent and dimensionless speciation rate $\nu$ are related even when
the latter is very small, implying that an observed value of the
exponent $z$ would predict the rate of introduction of new species
$\nu$.  Remarkably, the existence of positive correlations of these
two quantities is consistent with observational results. As an
example, it is known that close to the equator species-area exponents
tend to increase \citep{drakare2006} together with speciation rates
\citep{allen2006} and overall biodiversity \citep{stevens1986}.

In order to test the ecological plausibility of the relation between
$z$ and $\nu$, we make use of the definition of $\nu$ as the ratio
between the speciation rate $\sigma$ and the death rate $d$.  From
Eq.~(\ref{ourfit}) and separating the contribution from the variation
in the speciation rate from that of the variation in the death rate,
we have
\begin{equation}\label{comparisonequation}
\frac{1}{z}=q+m \ln(\nu)=q+m [\ln(\sigma)-\ln(d)]\,,
\end{equation}
where the arguments of the logarithm are made dimensionless by
measuring them in the same units. To ease the interpretation, we
recast this equality in the time domain using the lifespans $t=1/d$ and
the average time between speciation events $t^{(s)}=1/\sigma$:
\begin{equation}\label{comparisontimes}
\frac{1}{z}=q+m [\ln(t)-\ln(t^{(s)})]\,,
\end{equation}
The first term on the right hand side accounts for the variation in
$z$ due to the lifespan which is, of course, much easier to estimate
than the term due to speciation time and can still be important and
informative. Indeed there are evidences that taxa having a shorter
generation time have generally lower species area exponents
\citep{horner2004,green2004,zhou2008} (we recall that $m$ is
negative). We thus study how the inverse exponent $1/z$ varies with
the logarithm of the lifespan.  The results of this analysis are
presented in figure~\ref{figuradata} for data obtained from the
literature (see Appendix~\ref{appb} for a description of how data have
been collected), which shows that a linear relationship fits rather
well the data, with an observed slope $m_{meas}=-1.76\pm0.13$ (dashed
line in the figure) which is different from $m\approx -0.72$ predicted
by the voter model.

The fact that for species-area exponents measured in field data we
found $1/z\propto \ln(t)$ suggests a scaling relationship between
speciation time and lifespan, i.e.  that $t^{(s)}\sim t^{\gamma}$, so
that $m_{meas}= m(1-\gamma)$, as clear by substitution in the previous
formula.  We do not have any \textit{a priori} explanation for
justifying a power law dependence of the speciation time on the
lifespan, apart from the observation that the variations of many
ecologically relevant rates among species are governed by scaling laws
\citep{brown2004}.  We are not aware of independent estimation of the
dependence of the speciation time on the lifespan so to confirm or
reject the outcome of our analysis. 

However, the relation $m_{meas}=m(1-\gamma)$, with $m$ and $m_{meas}$
fixed by the voter model and field data respectively, yields a
negative $\gamma$.  This result is in contrast with biological
expectations as it would imply, e.g., a speciation time for bacteria
much longer than the one for trees, which is hard to justify
biologically.  Reasonable expectations would have been
$0<\gamma<1$. The limiting case of $\gamma=1$ is the trivial case in
which speciation time is proportional to the lifespan. This would have lead
to $m_{meas}=0$, i.e. same $\nu=t/t^{(s)}$ and $z$ for all taxa.  The
other limiting case is $\gamma=0$, which is plausible when the
possibility of creating a new species is triggered by some external
mechanism, like the availability of new niches, which is not strongly
correlated to any particular feature of the species. Another
justification could come from co-evolutionary mechanisms: species
having very different lifespans can still evolve on similar timescales
due to their ecological interactions \citep{thompson1992}. Actually,
co-speciation is known to occur in some cases of host-parasite systems
\citep{clayton2003}.  In this case, one would find $m_{meas}=m$.

\begin{figure}[t!]
\begin{center}
  \includegraphics[width=0.7\textwidth]{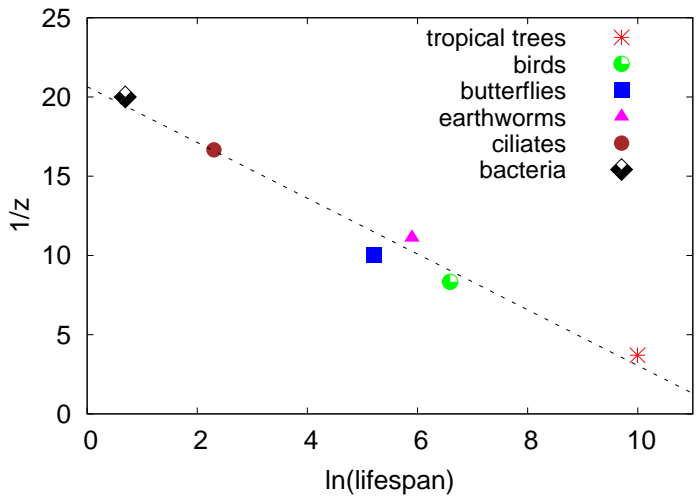}
  \caption{$1/z$ vs logarithm of lifespans (measured in days) for
    several taxa.  Dashed line is the best fit, giving a slope
    $m_{meas}=-1.76 \pm 0.13$ (reduced
    $\chi^2=1.04$).  \label{figuradata}}
\end{center}
\end{figure}

The inconsistent value of $\gamma$ we obtained can be interpreted
either as a failure of the basic assumptions of the neutral model and
thus of its inadequacy in describing realistic ecosystems or as the
need of additional mechanisms to be included in the framework of
dispersal models.  The quality of the linear fit shown in
Fig.~\ref{figuradata} would suggest to opt for the second
interpretation, even if the linear dependence of $1/z$ on $\log(t)$
needs to be tested with further measures.  In recent years, several
attempts of relaxing the strong assumptions of neutral models have
been tried.  The results of these models are pretty robust with
respect to modifications of some hypothesis such as the saturation of
the resources \citep{etienneb2007}. In spatial models, it has been
also observed that the introduction of trade-offs does not have a
dramatic effect on species-area exponents \citep{chave2002}.
Therefore, it is reasonable to search for other elements in the model
which can lead to a failure in reproducing the observed data. In
particular, the assumptions of a point speciation mode (i.e. the fact
that each individual has a fixed probability to speciate) is known to
be crucial and the results may change dramatically when considering
``fission modes'', corresponding to allopatric speciation
\citep{hubbell2001,hubbell2003,ricklefs2003,etienne2007}. Indeed, it
seems like the most important assumption of neutral models is that new
species enter the system with a population of a single individual
\citep{zillio2007}.  This could explain why speciation rates predicted
by neutral models with point speciation may look unrealistically high:
new species are introduced with only one individual and they have an
high probability of going extinct before being able to grow. In other
words, there could be a discrepancy between the parameter $\nu$ in the
model and the experimentally observed speciation rate. The effect of
realistic speciation mechanisms on neutral and more general dispersal
models could be key to understand the puzzle of the observed variation
of the exponents among different taxa.

In conclusion, simulations of the multi-type voter model for low
values of the speciation rate show a clear logarithmic functional
dependence of the specie-area exponent on the speciation rate and
independence on the dispersal kernel (provided it is not too short
ranged). Analysis of field data support a logarithmic dependence of
the exponent $z$ on the timescales of the problem, though with a
prefactor which is incompatible with that found in the model.  Our
analysis points out that more refined models should allow larger
variations in the exponent $z$ in order to be consistent with
observational data.

\section{Acknowledgments}
We would like to thank K. H. Andersen and F. Cecconi for a critical
reading of the manuscript.

\appendix

\section{Simulations and data analysis}
\label{appa}
By means of the dual representation of the voter model, the model
becomes equivalent to a system of coalescing random walkers, where a
birth/dispersal event corresponds to collision-coalescence of two
walkers and a speciation event to the death of a walker. Therefore, as
time proceeds, less and less surviving walkers should be accounted,
speeding up the simulation.  This allowed us to simulate lattices
$L\times L$ with $L=1024,2048$ and $L=4096$ for $\nu\in
[10^{-11}:10^{-3}]$. In order to test the effect of the system size
when boundaries are present, we have also used $L=512$; in this case
we generalized the algorithm by just refusing all moves causing the
exit of a walker from the simulation domain, thus constraining the
walkers to remain inside the initial grid. To embank the simulation
bottleneck due to the initial presence of $L^2$ walkers we optimized
the walker collision detection by means of a look-up table. For each
$\nu$ and $L$ we repeated the simulation many times with different
seeds for the random number generator, typically from $100$ to
$150-300$ for $L=2048,4096$ and $L=1024, 512$, respectively.  For the lower
values of $\nu$ simulations get very slow and a lower number of
realization was used, typically from $20$ to $60$ for $\nu<10^{-10}$.
Once the species occupancy patterns are obtained SAR curves are
derived by averaging the number of species in non-overlapping squares
of side $\sqrt{A}=1,\ldots, L$ whose union completely covers the
simulation grid. So that averages are performed both over the number
of sampled areas in each realization and over different realizations.
Statistical errors on the average are also computed.  The exponent $z$
characterizing the power law growth of $S$ with $A$ is then estimated
fitting by a linear least square method the function $\ln S= q+z\ln A$
with $A\in [A_{min}:A_{max}]$ chosen at the beginning and the end of
the intermediate regime, respectively.  The fit were performed by
minimizing the reduced $\chi^2$ function (i.e. normalizing the
$\chi^2$ with the number of degrees of freedom) but constraining the
minimal number of points to be considered (from $5$ to $15$ depending
on the extension of the intermediate range). As the least square error
is smaller than the variability of the fitted $z$ at changing the
minimal number of constrained points, we set the error on the estimate
as such variability. In Fig.~\ref{figuraexponents} errors are
comparable with the symbol size.

The quality of the fit is then compared
by a direct inspection of the local slopes (logarithmic derivatives,
i.e. $d\ln S/d\ln A$ vs $\ln A$) of the SAR curves.

\section{Details on observational data}
\label{appb}
 Data presented in
Fig.~\ref{figuradata} are based on the collection of exponents
presented by \citet{horner2004}. We avoided presenting the
$z$-value for plants since it varies a lot among different studies
(see supplementary information of \citet{horner2004}). As a
representative of long-lived organisms we have chosen tropical forest
trees which are well studied and we assumed for them $z=0.27$
\citep{lonsdale1999} and average lifespan $\approx 60$ years
\citep{condit1999}. The exponent $z$ values for butterflies,
earthworms, birds and ciliates are the same of the original reference;
the value for bacteria $z=0.05$ is an average between the value
$z=0.04$ in \citet{horner2004} and the value $z=0.06$ in a more
recent study by \citet{zhou2008}.  As far as the other lifespans are
concerned, we must stress that it can vary much from species to
species and, in most cases, it is hard to find in the literature good
statistical studies. However, due to the logarithmic dependence, the
fit is not be so sensitive to errors in the estimates as far as the
order of magnitude is correct.  We assumed an average lifespan of $2$
years for birds (see, e.g., \citet{speakman2005}). Despite their short
average lifespan in their adult stage, butterflies usually have a few
generations per year \citep{gilbert1975}; we assumed an average of two,
corresponding to $t=0.5$ years . Other estimated lifespans are $1$
year for earthworms \citep{earthworms}, $10$ days for ciliates
\citep{jensen2004} and $2$ days for bacteria \citep{clarholm1980}.

\end{document}